\documentclass[a4paper]{article}
\usepackage{amsmath,graphicx,hhline,microtype}
\usepackage{booktabs} 
\usepackage{multirow}
\usepackage{amssymb} 
\usepackage{pifont} 
\usepackage{url}
\usepackage{INTERSPEECH2021}

\newcommand{\xmark}{\ding{55}}%

\title{The Effectiveness of Time Stretching for Enhancing Dysarthric Speech for Improved Dysarthric Speech Recognition}

\name{Luke Prananta$^1$, Bence Mark Halpern$^{1,2,3}$, Siyuan Feng$^1$, Odette Scharenborg$^1$}
\address{
  $^1$Multimedia Computing Group, Delft University of Technology, the Netherlands\\
  $^2$Netherlands Cancer Institute, Amsterdam, the Netherlands\\
  $^3$ACLC, University of Amsterdam, Amsterdam, the Netherlands}
\email{\{S.Feng,O.E.Scharenborg\}@tudelft.nl, B.M.Halpern@uva.nl}

\begin{document}
\ninept
\maketitle
\begin{abstract}

In this paper, we investigate several existing and a new state-of-the-art generative adversarial network-based (GAN) voice conversion method for enhancing dysarthric speech for improved dysarthric speech recognition. We compare key components of existing methods as part of a rigorous ablation study to find the most effective solution to improve dysarthric speech recognition. We find that straightforward signal processing methods such as stationary noise removal and vocoder-based time stretching  lead to dysarthric speech recognition results comparable to those obtained when using state-of-the-art GAN-based voice conversion methods as measured using a phoneme recognition task. Additionally, our proposed solution of a combination of  MaskCycleGAN-VC and time stretched enhancement  is able to improve the phoneme recognition results for certain dysarthric speakers compared to our time stretched baseline.

\end{abstract}
\noindent\textbf{Index Terms}: dysarthric speech, voice conversion, dysarthric speech recognition, time stretching, generative adversarial networks

\section{Introduction}
\label{sec:intro}

Dysarthria is an encapsulating term for various motor speech disorders in which the muscles that produce speech are weakened or damaged. It is often the by-effect of degenerative diseases such as Parkinson’s disease or amyotrophic lateral sclerosis (ALS), but can also be caused by traumatic brain injuries or strokes. The reduced motor capabilities of certain speech muscles result in speech that is slurred and less intelligible. Dysarthria can greatly reduce a person’s quality of life and independence. Operating home appliances through voice could greatly improve these people's lives; however, dysarthric speech recognition performance is not good enough yet for practical applications, which means that there is a great need for high performance dysarthric speech recognition \cite{moro2019study}.

Dysarthric speech recognition is usually tackled from one of two perspectives, namely, data augmentation and dysarthric speech enhancement. The aim of data augmentation is to improve the recognition performance of dysarthric speech by training automatic speech recognition (ASR) models with synthetic dysarthric data. Previous approaches have tried to generate dysarthric data using different neural models including Transformers \cite{Harvill2021} and generative adversarial networks (GANs) \cite{Jiao2018}.

For dysarthric speech enhancement, voice conversion (VC) has been the dominant approach, although there are previous studies using time stretching, and formant synthesis \cite{rudzicz2013adjusting}. VC’s goal is to convert a source speaker’s speech (here: dysarthric speech) to a target speaker’s speech (here: healthy speech), while simultaneously retaining the linguistic content of the utterance. A VC task is either categorised as parallel or non-parallel. For parallel VC, the same utterances (containing the same linguistic content) are available from the source and the target speakers during training. When parallel utterances are not available, the VC task is non-parallel. Previous studies of dysarthric VC (DVC) have largely consisted of partial least squares regression- (PLS) \cite{aihara2017phoneme}, Gaussian mixture model (GMM) \cite{rudzicz2013adjusting}, or deep neural network-based (DNN) \cite{chen2020enhancing,parrotron} parallel methods. There are also some methods that incorporate non-parallel VC methods as part of a parallel VC system \cite{dvc-vtn-vae, matsubara2021high}.

Despite parallel models being able to synthesise highly natural speech using low amounts of data, the requirement of parallel data is a  limitation of VC methods, since it is substantially more difficult to collect than non-parallel data. 
Frameworks allowing non-parallel training thus have higher practical use. It is therefore not surprising that non-parallel CycleGAN-based approaches recently attracted some attention for DVC \cite{Purohit2020}. However, it is unclear what variant of CycleGAN-VC is the most ideal for DVC as \cite{Parmer_DiscoCycleGAN} found the DiscoGAN and CycleGAN-VC architectures comparable on objective metrics, while Mask-CycleGAN-VC \cite{kaneko2021maskcyclegan} seems to be better than CycleGAN-VC. Our study not only serves to fill this gap in the understanding of CycleGAN-based VC but importantly, also investigates  the efficiency of GAN-based VC methods compared to time stretching methods. Previous research has shown that the performance of plain time stretching is comparable to GMM-based parallel voice conversion on the measure of phoneme accuracy \cite{rudzicz2013adjusting}; however, it is unclear how it compares to state-of-the-art GAN-based methods.

In this paper, we investigate the efficiency of GAN-based methods for dysarthric-to-normal speech conversion with the aim to enhance dysarthric speech for improved dysarthric speech recognition, and compare their performance to that when using time stretching. We investigate state-of-the-art solutions for DVC using CycleGAN-based models in the form of an ablation study. Our main research questions are as follows: (\textbf{RQ1}) What aspects of CycleGAN-based non-parallel techniques are essential for enhancing dysarthric speech as indicated by improved dysarthric speech recognition performance (measured using the phoneme error rate; PER) (\textbf{RQ2}) How does the performance of state-of-the-art GAN-based methods for dysarthric-to-normal speech conversion compare to the performance when using time stretching?

\section{Methodology}
\label{sec:methodology}
\begin{table}[t]
\centering
\caption{Comparison of the key differences between the main GAN-based models tested in this paper, and the newly implemented intermediate models for the ablation study. 2-STEP stands for two-step adversarial loss, DTW stands for dynamic time warping, FIF DA stands for fill in the frame data augmentation.}
\label{tab:experiment_difference}
\resizebox{\columnwidth}{!}{%
\begin{tabular}{lccccc}
\toprule
 & Loss & Vocoder & 2-STEP  & DTW & FIF DA  \\
\midrule
CycleGAN-VC \cite{cycleganvc1} & $L_{1}$ & WORLD \cite{morise2016world} & \xmark & \xmark & \xmark  \\
DiscoGAN \cite{Purohit2020} & $L_{2}$ & AHOCODER \cite{erro2013harmonics} & \xmark & \checkmark & \xmark  \\
CycleGAN-VC + DTW  & $L_{1}$ & WORLD \cite{morise2016world} & \xmark & \checkmark & \xmark  \\
CycleGAN-VC + 2-STEP  & $L_{1}$ & WORLD  & \checkmark & \xmark & \xmark  \\
CycleGAN-VC + DTW + 2-STEP  & $L_{1}$ & WORLD & \checkmark & \checkmark & \xmark  \\
MaskCycleGAN-VC \cite{kaneko2021maskcyclegan} & $L_{1}$ & MelGAN \cite{kumar2019melgan} & \checkmark & \xmark & \checkmark  \\
\bottomrule
\end{tabular}%
}
\end{table}

\subsection{Dataset and conversion setup}
\label{subsec:dataset}

We used the UASpeech dataset \cite{uaspeech2008}, which contains parallel word recordings of 15 dysarthric speakers and 13 normal control speakers. Each speaker produced 455 utterances.  The subjective speech intelligibility of each speaker was judged by 5 non-expert American English native speakers. 

For our experiments, we followed the same data and speaker split as used in \cite{Purohit2020}, which is the following: During training, the dysarthric speech from four male speakers (M05, M08, M09 and M10) and four female speakers (F02, F03, F04 and F05) is used as source speech, while the healthy speech from four male healthy control speakers (CM05, CM08, CM09, CM10) and four female healthy control speakers (CF02, CF03, CF04 CF05) is used as target speech. A leave-one-out cross-validation scheme is used  for training and evaluating the models. Each model is trained with 1365 utterances from three different speakers, and evaluated with the 455 utterances from the remaining speaker.

\subsection{Preprocessing and denoising}
\label{subsec:preprocessing}

The speech data was denoised using a Python package called noisereduce \cite{tim_sainburg_2019_3243139}, which performs stationary noise removal based on the first 0.5 s of each utterances. Please note that this denoising step is by default applied on the UASpeech dataset since December 2020. Moreover, the paper \cite{Purohit2020} was written before December 2020 and does not mention the use of a denoising preprocessing step, we therefore believe that the difference in applying the denoising step is the most important difference between our work and the work in \cite{Purohit2020}.

Preliminary experiments showed that the quality of converted utterances are sensitive to the amount of silence in the audio. Therefore, we also performed an energy-based silence trimming (30 dB) using librosa \cite{mcfee2015librosa}. Moreover, we removed clicks from the audio using a simple heuristic (removing leading and trailing 0.2 s of the signal).

\subsection{Experimental design}

\subsubsection{Ablation study design}

In order to answer what aspects of the CycleGAN-based techniques are essential for enhancing dysarthric speech (\textbf{RQ1}), we compared the three CycleGAN-based models mentioned in the Introduction:  \textbf{DiscoGAN} (Section \ref{subsec:discogan}), \textbf{CycleGAN-VC} (Section \ref{subsec:cyclegan_vc}), and \textbf{MaskCycleGAN-VC} (Section \ref{subsec:mask_cyclegan_vc}). All models were trained using the data described in Section \ref{subsec:dataset}. ASR performance was measured in terms of the PER using the ASR specified in Section \ref{subsec:asr}.

It is important to recognise that the three GAN models are conceptually very similar, and that they consist of highly similar building blocks; however, important differences also exist between these models. Table \ref{tab:experiment_difference} gives an overview of the key differences between the GAN-based models tested in this work. These differences focus primarily on the use of a two-step adversarial loss (2-STEP), the use of dynamic time warping (DTW), and whether frame data augmentation (FIF DA) is used (explained in Section \ref{subsec:mask_cyclegan_vc}).
In order to pinpoint the importance of these aspects of the GAN models on dysarthric speech recognition performance, we carried out an ablation study for which we  created intermediate CycleGAN-based models where we added the 2-STEP loss and/or DTW to the CycleGAN-VC models:
\begin{description}
\item[CycleGAN-VC + DTW] CycleGAN-VC with dynamic time warping (DTW is explained in Section \ref{subsec:phase_vocoder})
\item[CycleGAN-VC + 2-STEP] CycleGAN-VC with two-step adversarial loss (Two-step adversarial loss is explained in Section \ref{subsec:mask_cyclegan_vc}) 
\item[CycleGAN-VC + DTW + 2-STEP] CycleGAN-VC with parallel data, DTW and two-step adversarial loss
\end{description}

Finally, in order to investigate the role of time stretching (\textbf{RQ2}), we used all the models (CycleGAN-VC, CycleGAN-VC + DTW, CycleGAN-VC + 2-STEP, CycleGAN-VC + 2-STEP + DTW, MaskCycleGAN-VC) with time stretched speech as input without retraining the VC models. The method of stretching for the time stretched speech will be described in Section \ref{subsec:phase_vocoder}. We will denote these models with the shorthand \textbf{+ TS}, and we will refer to these as \textit{time stretched models}.

\subsection{GAN architectures}

\subsubsection{CycleGAN-VC}
\label{subsec:cyclegan_vc}

CycleGAN-based VC aims to convert acoustic features from domain $\mathbf{x} \in X$ to domain $\mathbf{y} \in Y$ using a neural network $F$ as a forward-generator ($X \rightarrow Y$), and $G$ as the backward-generator ($Y \rightarrow X$), and another set of neural networks $D_X$ and $D_Y$ as the discriminators. The generators and the discriminators are optimised with regards to three loss functions: an adversarial loss function, a cycle-consistent loss function and an identity loss function.

\textbf{Adversarial loss function:} The aim of the adversarial loss function is to incentivise $G$ to fool the discriminator $D$, while $D$ is optimised to learn the difference between the distribution of samples generated by $G$ (the "fakes", usually denoted by 0) and the distribution of real samples (the "reals", usually denoted by 1),
\begin{align*}
    \mathcal{L}_{\text{GAN}}(G, D, X, Y) &= \mathbb{E}_{\boldsymbol{y}\sim p_{\text{data}}(\boldsymbol{y})}[\log{(0 - D(\boldsymbol{y}))}] \\ & + \mathbb{E}_{\boldsymbol{x}\sim p_{\text{data}}(\boldsymbol{x})}[\log{(1 - D(G(\boldsymbol{x})))}].
\end{align*}

\textbf{Cycle-consistent loss function:} The intuition of this loss is to measure the similarity between a sample and the same sample mapped to another domain and back to the original domain. Thus the cycle-consistency loss aims to minimise the difference between a sample $\boldsymbol{x} \in X$ and $F(G(\boldsymbol{x}))$, and the difference between a sample $\boldsymbol{y} \in Y$ and $G(F(\boldsymbol{y}))$
\begin{align*}
    \mathcal{L}_{\text{cycle}}(G, F) &= \mathbb{E}_{\boldsymbol{x}\sim p_{\text{data}}(\boldsymbol{x})}[\lVert F(G(\boldsymbol{x}))-\boldsymbol{x}\rVert_1] \\ & + \mathbb{E}_{\boldsymbol{y}\sim p_{\text{data}}(\boldsymbol{y})}[\lVert G(F(\boldsymbol{y}))-\boldsymbol{y}\rVert_1].
\end{align*}

\textbf{Identity loss function:} While the cycle-consistency loss constrains the structure of the mapping, on its own it does not suffice for preserving linguistic information \cite{cycleganvc1}. Therefore an identity-mapping loss is used to improve preservation of linguistic information. This loss was recommended for the original CycleGAN as well, to preserve colour composition between the input and output images.
\begin{align*}
    \mathcal{L}_{\text{id}}(G, F) &= \mathbb{E}_{\boldsymbol{y}\sim p_{\text{data}}(\boldsymbol{y})}[\lVert G(\boldsymbol{y})-\boldsymbol{y}\rVert_1] \\ & +  \mathbb{E}_{\boldsymbol{x}\sim p_{\text{data}}(\boldsymbol{x})}[\lVert F(\boldsymbol{x})-\boldsymbol{x}\rVert_1].
\end{align*}

The complete CycleGAN loss is then defined as follows:
\begin{align*}
    \mathcal{L}_{\text{CycleGAN}}(G, F, D_X, D_Y, X, Y) &= \mathcal{L}_{\text{GAN}}(G, D_Y, X, Y) \\ &+ \mathcal{L}_{\text{GAN}}(F, D_X, Y, X) \\ &+ \lambda_{\text{cycle}}\mathcal{L}_{\text{cycle}}(G, F).
\end{align*}

Our implementation of the CycleGAN-VC is based on a PyTorch implementation of CycleGAN-VC by\footnote{\url{https://github.com/karkirowle/cyclegan_pytorch}}. The CycleGAN-VC is used as the baseline model. In our ablation study, we create modifications of this CycleGAN model to obtain answers for our research questions.

We set up our CycleGAN-VC model with a fixed set of hyper-parameters. The model configuration followed the configuration proposed in \cite{cycleganvc1}.  The model used the Mel-generalised cepstrum (MCEP) features provided by the the WORLD vocoder \cite{morise2016world}. The pitch ($F_0$) features were log-speaker normalised during the conversion and the aperiodicities (AP) are simply copied.

The network was trained using an Adam optimiser with $\beta_1 = 0.5$ and $\beta_2 = 0.999$, with an initial learning rate of 0.0002 for the generator and 0.0001 for the discriminator. After $2\times 10^5$ iterations, the learning rates linearly decay over $2\times 10^5$ iterations. $\lambda_{\text{cycle}} = 10$ and $\lambda_{\text{id}} = 5$ are set to regularise the cycle-consistency loss $\mathcal{L}_{\text{cycle}}$ and the identity loss $\mathcal{L}_{\text{id}}$. After $1 \times10^4$ iterations,  $\mathcal{L}_{\text{id}}$ is set to $0$. A batch size of 1 is used during the training procedure, and a segment of 128 frames is randomly selected from a training sample. The CycleGAN-VC model was trained for 1000 epochs.

\subsubsection{DiscoGAN}
\label{subsec:discogan}

DiscoGAN differs from CycleGAN-VC described in Section \ref{subsec:cyclegan_vc} in the following aspects: (1) application of dynamic time warping to temporally align the source acoustic features (see Section \ref{subsec:phase_vocoder}), (2) usage of the AHOCODER instead of the WORLD vocoder, (3) a modification of the cycle-consistency loss function, called the mean squared error (MSE) cycle-consistency loss.

\textbf{MSE cycle-consistency loss function:} The mean squared error cycle-consistency loss function mimics the cycle-consistent loss function, except that it uses the $L_1$ norm instead of the $L_2$ norm:
\begin{align*}
    \mathcal{L}_{\text{cycle}}(G, F) &= \mathbb{E}_{\boldsymbol{x}\sim p_{\text{data}}(\boldsymbol{x})}[\lVert F(G(\boldsymbol{x}))-\boldsymbol{x}\rVert_2] \\ & + \mathbb{E}_{\boldsymbol{y}\sim p_{\text{data}}(\boldsymbol{y})}[\lVert G(F(\boldsymbol{y}))-\boldsymbol{y}\rVert_2].
\end{align*}

\subsubsection{MaskCycleGAN-VC}
\label{subsec:mask_cyclegan_vc}

MaskCycleGAN-VC differs from CycleGAN-VC by using (1) a different (MelGAN) vocoder, and therefore also a different feature representation; (2) a two step-step adversarial loss function; and (3) a data augmentation strategy called fill-in-the-frame data augmentation (FAF DA).

\textbf{Two-step adversarial loss:} The most notable difference is the introduction of two-step adversarial loss. This loss is added to address the over-smoothing caused by the cycle-consistency loss. Additional discriminators $D'_X$ and $D'_Y$ are used for a second adversarial loss for bidirectionally converted features. The loss is defined as follows:
\begin{align*}
    \mathcal{L}_{\text{GAN}_2}(G, F, D', X) &= \mathbb{E}_{\boldsymbol{x}\sim p_{\text{data}}(\boldsymbol{x})}[\log{(0 - D'(\boldsymbol{x}))}] \\ &+ \mathbb{E}_{\boldsymbol{x}\sim p_{\text{data}}(\boldsymbol{x})}[\log{(1 - D'(F(G(\boldsymbol{x}))))}].
\end{align*}

\textbf{Fill-in-the-frame data augmentation:} This is a data augmentation technique which randomly sets a temporal region in the source mel-spectrogram to zero with a binary mask. This random masking conditions the MaskCycleGAN-VC on a secondary task, namely, filling in the missing frames, alongside the original conversion task. Therefore, this data augmentation technique can alternatively viewed as a multi-task learning technique. For the  MaskCycleGAN-VC \cite{kaneko2021maskcyclegan} implementation we used the implementation provided here\footnote{\url{https://github.com/GANtastic3/MaskCycleGAN-VC.git}} with the same parameters. This model has the latest improvements and is a state-of-the-art CycleGAN model. The MaskCycleGAN-VC model is by and large similarly configured as the CycleGAN-VC model. However, the learning rates are set to decay after only $1 \times 10^4$ iterations. Segments of only 64 frames are randomly selected from the training samples. The MaskCycleGAN-VC model is trained for 300 epochs.

\subsection{Dynamic time warping and time stretching}
\label{subsec:phase_vocoder}

We experimented with two different techniques to temporally align the source and the target utterances: dynamic time warping (DTW) and phase vocoder-based time stretching. DTW implements a time alignment of the MCEP features of the source and the target utterances such that optimal warping paths are founds between these representations. We used the most common $L_2$ based DTW implementation from librosa \cite{librosa2015}.

For time stretching, we used the phase vocoder-based method which is provided by librosa \cite{mcfee2015librosa}. Phase vocoding resamples the magnitude short-time Fourier transform of the speech signal by linear interpolation, while simultaneously adjusting for the change in the phase. This results in a magnitude spectrogram with a smaller or larger number of analysis frames, which corresponds to a contracted or a stretched speech signal, respectively. In our work, we always adjusted the dysarthric speech to the duration of the target (healthy) speaker's speech. 

\subsection{Evaluation: Phoneme recognition}
\label{subsec:asr}

The different models in the ablation study were evaluated on a phoneme recognition task. We used a pre-trained, HMM-based Kaldi ASR model with the same specifications as the one used in \cite{Purohit2020} for phoneme recognition.

The ASR was trained with the TIMIT dataset, which is an English read speech corpus specifically designed for acoustic-phonetic studies \cite{garofolo1993darpa}. The UASpeech database does not come with phonemic transcriptions. We created these reference phoneme transcription using g2p-en\footnote{\url{https://github.com/Kyubyong/g2p}}, a tool for grapheme-to-phoneme conversion. In order to calculate the PER, the reference phoneme transcriptions are compared to the phoneme sequences predicted by the trained ASR for the VC utterances created by the different models.

\section{Results and discussion}

\begin{table*}[t]
\centering
\caption{An overview of the ASR performance in PER for all models. The percentages in parentheses indicate the subjective speech intelligibility taken from the UASpeech database. `DTW' denotes the models with a parallel data setup and DTW.  `2-STEP' denotes the models with second adversarial losses. `TS' denotes the models where time stretched dysarthric speech data is used as input. 'DN' denotes denoised dysarthric speech data is used as input. The results are separated into four blocks, which represent different sets of models: from top-to-bottom: models with only noise enhancement and no voice conversion that serve as the reference to compare to (GT); models trained on data that is enhanced using VC that do not use time stretching (DN $\emptyset$ TS); models trained on data that is enhanced using VC and  that use time stretched dysarthric speech as input (DN TS); and the models proposed by Purohit, these results are taken from \cite{Purohit2020} (P). \textbf{Bold} highlights column-wise the best result within a block, except for the first block where the rows serve as reference.}
\label{tab:result-table}
\resizebox{\textwidth}{!}{%
\begin{tabular}{ll|cccc|c|cccc|c}
\toprule
 & Model & M05 (58\%) & M08 (93\%) & M09 (86\%) & M10 (93\%) & Average (82.5\%) & F02 (29\%) & F03 (6\%) & F04 (62\%) & F05 (95\%) & Average (48\%)   \\
\midrule
\multirow{2}{*}{\rotatebox{90}{GT}} & Control (Healthy)     & 47.9\%  & 41.3\% & 51.9\% & 50.9\% & 48.0\%   & 51.98\%  & 57.2\% & 71.5\% & 46.7\%  & 56.8\%  \\ 
& Dysarthric                       & 96.1\%   & 60.2\% & 66.1\% & 64.6\% & 71.8\%    & 112.2\%   & 89.3\% & 78.0\% & 84.7\%  & 91.2\%  \\
\midrule
\multirow{5}{*}{\rotatebox{90}{DN $\emptyset$ TS}}  & CycleGAN-VC                & 110.4\% & \textbf{69.8\%} & 72.7\% & 80.5\% & 83.3\%   & 131.1\% & 103.6\% & 89.3\% & 100.0\% & 106.1\%  \\ 
& CycleGAN-VC + DTW          & 108.7\% & 74.0\% & \textbf{72.7\%} & 76.0\% & 82.8\%   & 111.0\% & 105.6\% & 84.1\% & 111.2\% & 103.1\%  \\ 
& CycleGAN-VC + 2-STEP       & 110.7\% & 73.1\% & 74.1\% & 77.8\% & 84.0\%  & 136.2\% & 103.8\% & 86.8\% & 100.0\% & 106.9\%  \\
& CycleGAN-VC + 2-STEP + DTW & 114.0\% & 74.0\% & 77.9\% & 78.7\% & 86.2\% & 132.2\% & 107.3\% & 96.0\% & 103.4\% & 109.8\%  \\
& MaskCycleGAN-VC & \textbf{105.7\%} & 71.4\% & 74.1\% & \textbf{62.9\%} & \textbf{78.5\%}   & \textbf{119.1\%} & \textbf{97.1\%} & \textbf{75.5\%} & \textbf{88.8\%}  & \textbf{95.3\%}  \\

\midrule

\multirow{6}{*}{\rotatebox{90}{DN TS}} & Dysarthric + TS                  & 73.8\% & \textbf{64.8\%}& \textbf{66.4\%} & \textbf{60.7\%} & \textbf{66.4\%}   & 80.4\% & \textbf{76.9\%} & 72.6\% & \textbf{63.7\%} & 73.4\%  \\ 

& CycleGAN-VC + TS                 & 76.8\% & 67.8\% & 72.3\% & 72.9\% & 72.4\%     & 80.2\% & 83.3\% & 79.3\% & 74.0\% & 79.2\% \\
& CycleGAN-VC + DTW + TS           & 75.3\% & 71.6\% & 71.3\% & 70.9\% & 72.3\%     & 81.2\% & 87.6\% & 77.7\% & 86.6\% & 83.3\%  \\ 
& CycleGAN-VC + 2-STEP + TS        & 78.7\% & 79.3\% & 77.8\% & 76.4\% & 78.1\%   & 80.8\% & 83.3\% & 79.2\% & 75.3\% & 79.7\%  \\
& CycleGAN-VC + 2-STEP  + DTW + TS & 76.7\% & 75.6\% & 78.0\% & 76.4\% & 76.7\% & 81.8\% & 86.6\% & 84.5\% & 77.6\% & 82.6\%  \\ 
& MaskCycleGAN-VC + TS             & \textbf{65.1\%} & 65.2\% & 71.7\% & 65.2\% & 66.8\%   & \textbf{72.4\%} & 81.0\% & \textbf{70.6\%} & 69.0\% & \textbf{73.2\%} \\ 
\midrule
\multirow{4}{*}{\rotatebox{90}{P \cite{Purohit2020}}} & Control (Healthy)  & - & - & - & - & 64.7\% & - & - & - & - & 65.4\% \\ 
& Dysarthric  & - & - & - & - & 77.9\% & - & - & - & - & 87.1\% \\ 
& DNN  & - & - & - & - & 82.9\%& - & - & - & - & 75.7\% \\ 
& DiscoGAN  & - & - & - & - & \textbf{73.3\%} & - & - & - & - & \textbf{71.1\%} \\ 
\bottomrule
\end{tabular}%
}
\end{table*}

\label{sec:discussion}

Table \ref{tab:result-table} gives an overview of the performance of all models.  The PER results are shown for individual speakers separately and averaged over all speakers and blocked by model type (see the table caption for an explanation).  
Note that empty cells refer to results that were not given or specified by \cite{Purohit2020}.

\subsection{RQ1: GAN modifications}
\label{5modifications}

To investigate what aspects of GAN models are important for enhancing dysarthric speech for improved dysarthric speech recognition, we first compare the results of the CycleGAN-VC baseline to the results of the newly created, intermediate CycleGAN-based models in the second block of Table \ref{tab:result-table} (DN $\emptyset$ TS). 

Comparing the row CycleGAN-VC with the other rows in the second block (DN $\emptyset$ TS) shows us that the addition of DTW (CycleGAN-VC+DTW) results in 0.5\% absolute improvement in the case of the male speakers and 3.0\% absolute improvement in the case of female speakers. The better performance of the CycleGAN-VC + DTW indicates that the DTW seems to slightly improve the temporal aspects of the dysarthric speech.

Both intermediate models including the 2-STEP (CycleGAN-VC + 2-STEP + (DTW)) have a performance that is on average worse compared to the CycleGAN-VC model. We hypothesise that the over-smoothing issue that the 2-STEP is supposed to alleviate might be essential for the naturalness of the synthesised speech, but not for enhancing the speech signal for improved dysarthric speech recognition.

The MaskCycleGAN-VC variant, which uses 2-STEP and FIF DA, is the best performing GAN-variant, with a 4.8\% absolute improvement in the case of the male speakers, and 10.8\% absolute improvement in the case of the female speakers. So despite incorporating the 2-STEP, the MaskCycleGAN-VC variant has the best performance.

The better performance for the MaskCycleGAN-VC can be explained by the following reasons. First, an improved MelGAN vocoder is used instead of the WORLD vocoder, which also means that higher dimensional mel-spectrogram features are used instead of MCEPs. Secondly, the fill-in-the-frame data augmentation technique is used, increasing the amount of available data for training. We suspect that these two aspects are jointly responsible for the MaskCycleGAN-VC outperforming the other models but this question is left for future research.  

To summarise, we observe that the different GAN architectures lead to relatively similar performances of enhanced dysarthric speech recognition, meaning that the different aspects  do not have a major effect on the intelligibility of the converted speech as measured by the PER. The slightly better results for MaskCycleGAN-VC and CycleGAN-VC + DTW suggest that the most important aspects for the success of GAN models are the choice of vocoder (MelGAN), the FAF DA, and the application of DTW.

\subsection{RQ2: Effectiveness of time stretching}
\label{subsection:effectiveness}

To examine the effectiveness of time stretching, we first compared the recognition performance of the different models using time stretched speech (DN TS; third block of Table \ref{tab:result-table}) to that of the different models without time stretched speech (DN $\emptyset$ TS; second block of Table \ref{tab:result-table}). This will be followed by a within-block comparison of the models using time stretched speech and 
a comparison of these models to time stretched dysarthric (Dysarthric + TS; block 3) speech.

First, we observe that all of the time stretched GAN-based models outperform their non-time stretched counterparts. The improvements are substantial, ranging from 5.9\% (for CycleGAN-VC + 2-STEP) to above 9.0\% (all other models). 

Second, the comparison of the different time stretched models shows that the order of performance for the models is mostly consistent with the order of the performance of the models that do not use time stretched speech: The MaskCycleGAN-VC still outperforms the CycleGAN-VC baseline, however the improvements introduced by the DTW seem to be either marginal (in the case of males) or specific to certain speakers (in the case of females). The DTW and the time stretching both aim to improve the temporal aspects of the speech. We hypothesise that after time stretching the DTW becomes unnecessary, which explains the smaller improvement in recognition performance for DTW for time stretched dysarthric speech.

Third, and most surprisingly, we observe that simple time stretching of the dysarthric speech on average outperforms the best performing GAN-model (MaskCycleGAN-VC + TS), with the exception of speakers M05, F02, and F05, which are mid to high severity cases. Therefore, we conclude that time stretching speech seems to be a better solution for improving dysarthric speech recognition than using purely GAN-based methods for the low and very high severity cases. Nevertheless, GAN-based methods can complement time stretching to improve the spectral structure of the dysarthric speech (\textbf{RQ2}) for the mid severity cases. 

\subsection{Comparison of the results to state-of-the-art and DiscoGAN}
\label{section:comparison_to_purohit}

Our results cannot be directly compared to those reported in \cite{Purohit2020} and their state-of-the-art DiscoGAN model because of the differences in the denoising of the dysarthric speech that was applied in our work but not in theirs (see Section \ref{subsec:preprocessing}).

The comparison of our ground truth (GT) healthy speech and dysarthric speech (block 1) and Purohit's results (P) (block 4) in Table \ref{tab:result-table} shows the impact of the denoising of the dysarthric speech on recognition performance: we observe overall a lower, thus better, average PER for the denoised speech for healthy speech and for dysarthric speech from male speakers. It is especially important to point that our dysarthric male speech result is already better than Purohit's proposed DiscoGAN model.

The CycleGAN-VC - while being conceptually similar to the DiscoGAN -  peforms worse than DiscoGAN.  The observed differences are most likely due to the differences in the CycleGAN-VC and DiscoGAN setup (see also Table \ref{tab:experiment_difference}): (1) the AHOCODER \cite{ahocoder2011} vocoder is used in \cite{Purohit2020}, while we use the WORLD vocoder, (2) CycleGAN uses a single cycle-consistency loss while DiscoGAN uses individual reconstruction losses. However, these losses are conceptually similar, therefore we expect (2) to have only a minor influence on the results compared to the vocoder.

\section{Conclusions}
\label{sec:conclusions}

We investigated several existing and a new state-of-the-art generative adversarial network-based (GAN) voice conversion methods for enhancing dysarthric speech for improved dysarthric speech recognition. Our main finding is that time stretching dysarthric speech to the target (healthy) speaker's rate improves dysarthric speech recognition performance to a level that is comparable, and even outperforms that of, existing state-of-the-art generative adversarial networks. The application of MaskCycleGAN-based voice conversion on time stretched speech yields results that are slightly better than pure time stretching, but only for mid to high severity speakers with dysarthria. The current performance of dysarthric speech enhancement is unfortunately still not good enough for practical use. The enhancement results proved to be highly dependent on the temporal structure of the speech, as demonstrated by the improved performance of the models using time stretching (+ TS) and dynamic time warping (+ DTW). Therefore, future work should focus on sequence-to-sequence based architecture, which have  already shown to excel at improving the temporal structure of dysarthric speech for parallel voice conversion \cite{N2D_VC}.

\section{Acknowledgements}
This project has received funding from the EU’s H2020 research and innovation programme under MMSC grant agreement No 766287. The Department of Head and Neck Oncology and surgery of the Netherlands Cancer Institute receives a research grant from Atos Medical (H\"orby, Sweden), which contributes to the existing infrastructure for quality of life research.

\bibliographystyle{IEEEtran}
\bibliography{strings,refs,thesis}

\end{document}